\documentclass[showpacs,showkeys,12pt,preprint,preprintnumbers,nofootinbib,groupedaddress,superscriptaddress,amsmath,amssymb]{revtex4}

\usepackage{graphicx}
\usepackage{dcolumn}
\usepackage{bm}
\usepackage{amsmath}
\usepackage{epsfig} 

\def\beq{\begin{equation}}
\def\eeq{\end{equation}}
\def\bea{\begin{array}}
\def\eea{\end{array}}
\def\be{\begin{equation}}
\def\ee{\end{equation}}
\def\ba{\begin{eqnarray}}
\def\ea{\end{eqnarray}}

\def\to{\rightarrow}

\def\[{\left[}
\def\]{\right]}
\def\({\left(}
\def\){\right)}

\def\sm0{{\widetilde{m}_0}}

\def\U1em{{U(1)_{\rm em}}}
\def\to{\rightarrow}

\def\sq2{\sqrt{2}}

\def\ee{e^+e^-}

\def\End{\end{document}}

\newcommand{\gsim}{\mbox{ \raisebox{-1.0ex}{$\stackrel{\textstyle >}
{\textstyle \sim}$ }}}
\newcommand{\lsim}{\mbox{ \raisebox{-1.0ex}{$\stackrel{\textstyle <}
{\textstyle \sim}$ }}}

\def\fsl#1{\setbox0=\hbox{$#1$}                 
   \dimen0=\wd0                                 
   \setbox1=\hbox{/} \dimen1=\wd1               
   \ifdim\dimen0>\dimen1                        
      \rlap{\hbox to \dimen0{\hfil/\hfil}}      
      #1                                        
   \else                                        
      \rlap{\hbox to \dimen1{\hfil$#1$\hfil}}   
      /                                         
   \fi}

\begin{document}

\title{Probing the Majorana nature of TeV-scale radiative seesaw models at collider experiments}
\author{Mayumi Aoki}
\email{mayumi@tuhep.phys.tohoku.ac.jp}
\affiliation{Department~of~Physics,~Tohoku~University,~Aramaki,~Aoba,~Sendai,~Miyagi~980-8578,~Japan}
\author{Shinya Kanemura}
\email{kanemu@sci.u-toyama.ac.jp}
\affiliation{Department~of Physics,~University~of~Toyama,~3190~Gofuku,~Toyama~930-8555,~Japan}
%
\preprint{TU-853, UT-HET 030}
\pacs{\, 12.60.Fr, 14.60.St, 14.80.Fd} 
\keywords{\, Radiative seesaw models, Higgs physics, Flavor physics}

\begin{abstract}
 A general feature of TeV-scale radiative seesaw models, in
 which tiny neutrino masses are generated 
 via loop corrections, is an extended scalar (Higgs) sector.
 Another feature is the Majorana nature; {\it e.g.}, introducing
 right-handed neutrinos with TeV-scale Majorana masses under the discrete symmetry, or otherwise introducing some
 lepton number violating interactions in the scalar sector.  
 We study phenomenological aspects of these models at collider experiments.
 We find that, while properties of the extended Higgs sector of these models 
 can be explored to some extent,
 the Majorana nature of the models can also be tested directly at the
 International Linear Collider via the electron-positron and
 electron-electron collision experiments.
\end{abstract}

\maketitle

\setcounter{footnote}{0}
\renewcommand{\thefootnote}{\arabic{footnote}}

\section{Introduction} 

The neutrino data show that neutrinos have tiny masses as
compared to the electroweak scale. 
This is clear evidence for physics beyond the standard model (SM).
The data also indicate that the structure of flavor mixing for neutrinos is largely different from that for charged leptons.
These facts would suggest that, while charged leptons have Dirac type
masses,  the neutrino masses are of the Majorana type.
The tiny Majorana masses of left-handed neutrinos are generated
from the dimension-five effective  operators
\begin{eqnarray}
   {\cal L} = \frac{c_{ij}}{2 \Lambda} \overline{\nu^c}_L^i {\nu}_L^j \phi^0 \phi^0, \label{eq:5op}
\end{eqnarray}
where $\Lambda$ represents a mass scale, $c_{ij}$ are dimensionless
coefficients, and $\phi^0$ is the Higgs boson.
After electroweak symmetry breaking, the mass matrix $M_\nu^{ij}$
for left-handed neutrinos appears as
   $M_\nu^{ij} = {c_{ij} \langle{\phi^0}\rangle^2}/{\Lambda}$. 
As the vacuum expectation value (VEV) $\langle \phi^0 \rangle$ of the Higgs
boson is ${\cal O}(100)$ GeV, the observed tiny neutrino masses
($M_\nu^{ij} \lsim 0.1$ eV) are realized when
$(c_{ij}/\Lambda) \sim {\cal O}(10^{-14})$ GeV$^{-1}$. 
It has been an interesting problem how we can naturally
explain such a small number with less fine tuning.%

If the operators in Eq.~(\ref{eq:5op}) appear at the tree level
in the low energy effective theory, $\Lambda$ has to be as large as
${\cal O}(10^{8})-{\cal O}(10^{14})$ GeV for $c_{ij}$ being ${\cal
O}(10^{-6})-{\cal O}(1)$ to describe the data.
For example, in the tree-level seesaw scenario where right-handed 
neutrinos are introduced, their Majorana masses have
to be set much higher than the electroweak scale~\cite{see-saw},
corresponding to the scale $\Lambda$ in Eq.~(\ref{eq:5op}).
Although the scenario is simple, it requires another hierarchy between
the mass of right-handed neutrinos and the electroweak scale, and in
addition, physics at such a large mass scale is difficult to
be tested at collider experiments.

Quantum generation of neutrino masses is an alternative way to
obtain $(c_{ij}/\Lambda) \sim {\cal O}(10^{-14})$ GeV$^{-1}$.
Due to the loop suppression factor, $\Lambda$ in these models  
can be lower  as compared to that in the tree-level seesaw models.
Consequently, the tiny neutrino masses would be explained in a
natural way by the TeV-scale dynamics without introducing very high
mass scales.  
The original model of this line was proposed by Zee~\cite{zee},
where neutrino masses were generated at the one-loop level.
Some variations were considered~\cite{radiative,ZeeBabu,knt,MaModel,aks-prl}, for example, 
by Zee and Babu~\cite{ZeeBabu},
Kraus-Nasri-Trodden~\cite{knt}, Ma~\cite{MaModel}, and
the model in Ref.~\cite{aks-prl}.
The last three models contain dark matter (DM) candidates with the odd
quantum number under the discrete $Z_2$ symmetry.
It must be a charming point in these TeV-scale radiative seesaw models that they are directly
testable at the collider experiments such as Large Hadron Collider (LHC)
and the International Linear Collider (ILC).

 A general feature in radiative seesaw models
 is an extended Higgs sector, whose
 detail is strongly model dependent.  
 The discovery of these extra Higgs bosons and detailed measurements of
 their properties at current and future collider experiments can give partial 
 evidence for the radiative seesaw models.  
 In the literature~\cite{
 radiative-ph,Babu:2002uu,AristizabalSierra:2006gb,Nebot:2007bc,Ma-DM,kingman-seto,aks-prd}, phenomenology of these radiative
 seesaw models has already been studied extensively.
 Such previous works mainly discuss constraints on the
 flavor structure from the current data for such as neutrino physics
 and DM, and also study collider phenomenology of 
 the Higgs sectors~\cite{Deshpande:1977rw,Barbieri:2006dq,CaoMa,LopezHonorez:2006gr,Goh:2009wg,Su:2009fz,
 typeX,Logan:2009uf,Belyaev:2009zd}.

Another common feature in radiative seesaw models is the Majorana nature.
 In order to induce tiny Majorana masses for left-handed neutrinos,
 we need to introduce its origin such as
 lepton number violating interactions in the scalar
 sector~\cite{zee,ZeeBabu} or right-handed neutrinos with TeV-scale
 Majorana masses~\cite{knt,MaModel,aks-prl}.
 When the future data would indicate an extended Higgs sector
 predicted by a specific radiative seesaw model,
 the direct detection of the Majorana property at collider experiments should be
 a fatal probe to identify the model.

 In this Letter, we study the phenomenology in TeV-scale radiative seesaw models,   
in particular, a possibility of detecting the Majorana nature at collider experiments.
 We mainly discuss three typical radiative seesaw models as
 reference models; the model by Zee and Babu where neutrino masses are
 generated at the two-loop level~\cite{ZeeBabu}, that by Ma with one-loop
 neutrino mass generation~\cite{MaModel},
 and that in Ref.~\cite{aks-prl}  where neutrino masses are
 generated at the three-loop level.
 Typical parameter regions where the data can be satisfied have
 been already studied in each model in the literature.
 We here study collider phenomenology in such typical parameter regions in
 each model, and discuss the discrimination of these models by measuring
 the details of the Higgs sector and the Majorana nature at the LHC and the ILC.

\begin{figure}[t]
\includegraphics[width=16.8cm,angle=0]{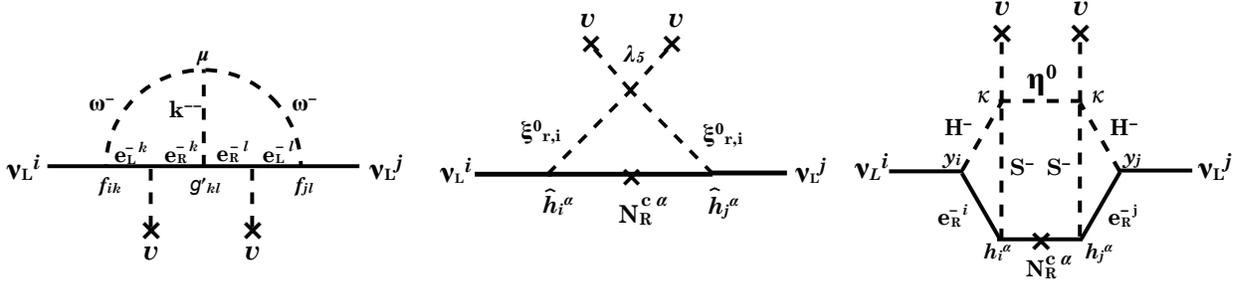}
  \caption{
 Feynman diagrams for neutrino masses in the model by Zee-Babu~\cite{ZeeBabu}
 (left), that by Ma~\cite{MaModel} (center) and that in Ref.~\cite{aks-prl} (right).
 }
  \label{fig:diag}
\end{figure}

\section{Radiative seesaw models}

\subsection{The Zee-Babu model}

In the model proposed in Ref.~\cite{ZeeBabu} (we refer to as the
Zee-Babu model), in addition to singly-charged singlet scalar bosons
$\omega^\pm$, doubly-charged singlet scalar fields $k^{\pm\pm}$ are
introduced,  both of which carry
the lepton number of two-unit, and their interactions are given by  
\begin{eqnarray}
 {\cal L}_{\rm int} =
f_{ab} (\overline{{L}_{aL}^{ci}} L_{bL}^{j}) \epsilon_{ij} \omega^+
+       g'_{ab} (\overline{{\ell}_{aR}^{c}} \ell_{bR}^{}) k^{++} 
- \mu \, 
  k^{++} \omega^- \omega^-
+ {\rm H.c.},
  \end{eqnarray}
where $L_L^{}$ is the left-handed lepton doublet and $\ell_R^{}$ is the right-handed lepton singlet.
The matrices $f_{ij}$ and $g'_{ab}$ are respectively an anti-symmetric and a symmetric couplings and 
the lepton number is violated by the interaction with the
parameter $\mu$.

The neutrino mass matrix is generated at the two-loop level via the 
diagram in Fig.~\ref{fig:diag} (left);
\begin{eqnarray}
   M_{ij}^\nu = \sum_{k,\ell=1}^3
   \left(\frac{1}{16 \pi^2}\right)^2 
  \frac{4\mu}{m_{\omega}^2} f_{ik}  (y_{\ell_{k}} g_{k\ell}
  y_{\ell_{\ell}}) f_{\ell j} v^2  I_1(m_k^2/m_\omega^2),  
 \end{eqnarray}
where $y_{i}$ [$=\sqrt{2} m_i/v$ ($i=e$, $\mu$, $\tau$)] are the SM Yukawa
coupling constants of charged leptons with the masses $m_i$ 
and the VEV $v$ ($\simeq 246$ GeV),
$g_{ij}$ are defined as $g_{ii}=g_{ii}'$ and $g_{ij}=2g_{ij}'$
($i\neq j$), $m_{\omega}$ and $m_{k}$ are masses of $\omega^{\pm}$ and
$k^{\pm\pm}$, and
\begin{eqnarray}
  I_1(r)= -\int_0^1dx \int_0^{1-x}dy \frac{1}{x+(r-1)y+y^2} \ln\frac{y(1-y)}{x+ry},
 \end{eqnarray}
where $I_1(r)$ takes the value of around 3 - 0.2 for $10^{-2} \lsim r \lsim 10^2$.
The universal scale of neutrino masses is determined by the two-loop
suppression factor $1/(16\pi^2)^2$ and the lepton number violating
parameter $\mu$. The charged lepton Yukawa coupling
constants $y_{\ell_{i}}$ ($y_e \ll y_\mu \ll y_\tau \lsim 10^{-2}$)
give an additional suppression factor.
Thus, any of $f_{ij}$ or $g_{ij}$ can be of ${\cal O}(1)$ when
$m_\omega$ and $m_k$ are at the TeV scale.
The flavor structure of the mass matrix is determined
by the combination of the coupling constants $f_{ij}$ and $y_{i} g_{ij} y_{j}$.

The flavor off-diagonal coupling constants $f_{ij}$ and $g_{ij}$ induce lepton flavor violation (LFV).
From the results of
$\mu\to e\gamma$, $\tau\to e\gamma$ and $\tau\to\mu\gamma$,
$|f_{\mu\tau}f_{\tau e}|$,  $|f_{\tau\mu}f_{\mu e}|$ and $|f_{\tau e}f_{e\mu}|$
are respectively constrained as a function of $m_{\omega}$. 
The data of rare decays of
$\mu\to eee$, $\tau\to \mu \mu e$ and $\tau\to \mu e e$
are also used to constrain the combinations $|g_{\mu e}g_{ee}|$,
$|g_{\tau e}g_{\mu \mu}+g_{\tau \mu}g_{\mu e}|$ and 
$|g_{\tau e}g_{\mu e}+g_{\tau \mu}g_{e e}|$,
respectively, depending on $m_{k}$.
The $g-2$ data can also be used to constrain a combination of
these coupling constants with $m_\omega$ and $m_{k}$\footnote{
If we take $g_{ee}=0$, then  $m_{k}$ is unbounded from the
 $\mu\to eee$ and $\tau\to \ell ee$ results ($\ell=e$ or $\mu$),
 so that relatively
 light $k^{\pm\pm}$ ($m_{k} \sim 100-200$ GeV) are possbile. }.

In the scenario with hierarchical neutrino masses,
$f_{ij}$ satisfy $f_{e\mu} \simeq f_{e\tau} \simeq f_{\mu\tau}/2$.
The typical relative magnitudes among the coupling constants $g_{ij}$
can be
$g_{\mu\mu}: g_{\mu\tau}: 
g_{\tau\tau} \simeq 1 :  m_\mu/m_\tau: (m_\mu/m_\tau)^2$. 
For $g_{\mu\mu}\simeq 1$, the neutrino data and the LFV data give 
the constraints such as $m_k \gsim 770$ GeV and $m_\omega \gsim 160$ GeV~\cite{AristizabalSierra:2006gb}. 
On the other hand, the constraints on the couplings and masses are more stringent for the inverted neutrino mass 
hierarchy. The current data then gives $m_\omega \simeq 825$ GeV for $g_{\mu\mu}\simeq 1$~\cite{AristizabalSierra:2006gb}. 
One of the notable things in this case is the lower bound on $\sin^22\theta_{13}$, which is predicted as around 
0.002~\cite{Babu:2002uu}. 
%

\subsection{Models with TeV-scale Right-handed Neutrinos with a discrete $Z_2$ symmetry}

Similar to the tree-level seesaw model, tiny masses of left-handed neutrinos 
would also come from Majorana masses $M_{N_R^\alpha}$ of gauge-singlet
right-handed neutrinos $N_R^\alpha$ in the radiative seesaw scenario~\cite{knt,MaModel,aks-prl}.    
One simple way to realize the absence of the tree-level Yukawa
interaction $\overline{\nu}_L^i \tilde{\Phi} N_R^\alpha$
is introduction of a discrete $Z_2$ symmetry,
with the assignment of the odd quantum number to $N_R^\alpha$ and the
even to the SM particles.
To obtain the dimension five operator in Eq.~(\ref{eq:5op}) at the loop level,
we need to introduce additional $Z_2$-odd scalar fields. 
The lightest of all the $Z_2$-odd particles can be a candidate of DM if it is
electrically neutral.
The original model of the radiative seesaw model with such a discrete symmetry
is proposed by Krauss, Nasri, and Trodden~\cite{knt}, in which neutrino masses are
induced at the three-loop level.
In the following, we consider two variant models of the Kraus-Nasri-Trodden (KNT) model.

\subsubsection{The Ma model}

The model in Ref.~\cite{MaModel}, which we here refer to as the Ma model,
is the simplest radiative seesaw model with right-handed neutrinos
$N_R^\alpha$, in which the 
discrete $Z_2$ symmetry is introduced and its odd quantum number is
assigned to $N_R^\alpha$.
The Higgs sector is composed of two Higgs doublet fields, one of which 
($\Xi$) is $Z_2$ odd. As long as the $Z_2$ symmetry is exact, the neutral
components of $\Xi$ do not receive VEVs.
We have one SM-like Higgs boson $h$,
and four physical $Z_2$-odd scalar states; $\xi_r^{0}$ (CP-even), $\xi_i^{0}$ (CP-odd)
and $\xi^\pm$ as physical scalar states.
This $Z_2$ odd Higgs doublet is sometimes called as the inert Higgs doublet~\cite{Deshpande:1977rw}
or the dark scalar doublet~\cite{CaoMa}.
 The LEP II limits are studied in this model in Ref.~\cite{IDM-LEPII}.

The neutrino masses are generated at the one loop level via the diagram
depicted in Fig.~\ref{fig:diag} (center),
in which $Z_2$ odd particles, $\xi^0$ and $N_R^\alpha$, are in the loop.
The mass matrix is calculated as
\begin{eqnarray}
  M_{ij}^\nu = -\sum_{\alpha=1}^3 \left(\frac{1}{16 \pi^2}\right) 
    \frac{\hat{h}_{i}^\alpha \hat{h}_j^\alpha  \lambda_5 v^2}{M_{N_{R}^\alpha}}   
      \frac{1}{1-r^\alpha} \left(1 + \frac{1}{1-r^\alpha}\ln r^\alpha \right), 
 \end{eqnarray}
 where $\hat{h}_i^\alpha$ are the Yukawa coupling constants of
 $\overline{\nu}_L^i \widetilde{\Xi} N_R^\alpha$, $M_{N_R^\alpha}$ is
 the Majorana mass of the $\alpha$-th generation right-handed neutrino
 $N_R^\alpha$, $\lambda_5=(m_{\xi_i^{}}^2-m_{\xi_r^{}}^2)/v^2$,
 $r^\alpha=m_0^2/M_{N_R^\alpha}^2$ with $m_0=(m_{\xi_i}^{}+m_{\xi_r}^{})/2$,
 where $m_{\xi_r}^{}$ and $m_{\xi_i}$ are masses of $\xi_r^0$ and $\xi_i^0$, respectively.
The universal scale for neutrino masses is determined by
the one-loop suppression factor $1/(16\pi^2)$,
$\lambda_5$ and $M_{N_R^{\alpha}}$.
The flavor structure in $M_{ij}^\nu$ is realized by the combination
of $\hat{h}_i^\alpha \hat{h}_j^\alpha/M_{N_R^\alpha}$.
Therefore, for $M_{N_R^\alpha} \sim {\cal O}(1)$ TeV, the combination of the coupling
constants would be 
$|\lambda_5| (\hat{h}_i^\alpha)^2 \sim 10^{-9}$.
 
In this model, there are two scenarios with respect to the DM candidate; i.e.,
the lightest right-handed neutrino $N_R^1$
or the lightest $Z_2$-odd neutral field ($\xi_r^{0}$ or $\xi_i^{0}$).
For both cases, there are parameter regions where the neutrino data are
adjustable without contradicting other phenomenological
constraints~\cite{Ma-DM}.
In this Letter, we consider the case where the dark doublet
component $\xi_r^{0}$ is the DM candidate\footnote{ In Ref.~\cite{Ma-DM}, the scenario where the lightest right-handed
neutrino is DM is explored. 
}.
When the mass of the DM is around 50 GeV,
the typical value of $\lambda_5 \sim 10^{-2}$ 
for the neutrino masses gives the mass difference between $\xi_r^{0}$ and $\xi_i^{0}$ about 10 GeV.\footnote{In order to avoid constraint from the DM direct search results, it is
required that $|\lambda_5| > 10^{-6}$.}
The relic abundance of such DM is consistent with the WMAP data~\cite{LopezHonorez:2006gr}.

\subsubsection{The AKS model}

In the model in Ref.~\cite{aks-prl}, which we here refer to as the AKS
model, it is intended that not only
the tiny neutrino masses and DM but also baryon asymmetry
of Universe are explained at the TeV scale.
In addition to the TeV-scale right-handed neutrinos $N_R^\alpha$ ($\alpha=1,2$),
the Higgs sector is composed of $Z_2$-even two Higgs doublets $\Phi_i$ ($i=1,2$)
and $Z_2$-odd charged singlets $S^\pm$ and a $Z_2$-odd neutral real singlet $\eta^0$.
Therefore the physical states in the $Z_2$-even sector are $H$ (CP-even), $A$ (CP-odd),
$H^\pm$ and $h$ (CP-even).

The neutrino mass matrix is generated at the three-loop level via 
the diagram in Fig.~\ref{fig:diag} (right), and is expressed as 
\begin{eqnarray}
 M_{ij}^\nu = \sum_{\alpha=1}^{2} \left(\frac{1}{16 \pi^2}\right)^3
      \frac{(y_{\ell_i}^{} h_i^\alpha) (y_{\ell_j}^{}
   h_j^\alpha) (\kappa \tan\beta)^2 v^2}{M_{N_R^\alpha}^{}}
     I_2(m_{H^\pm}^{},m_{S^\pm}^{},m_{N_R^\alpha},m_{\eta}),  
 \end{eqnarray}
where $m_{H^\pm}$, $m_{S^\pm}$, $m_{N_R^\alpha}$ and $m_{\eta}$ are
the masses of the doublet originated charged Higgs boson $H^\pm$,
$S^\pm$, $N_R^\alpha$ and $\eta^0$, respectively; 
$h_i^\alpha$ and $\kappa v$ are the coupling constants of $\overline{N}_{R}^\alpha e_R^i
S^+$ and $H^+S^-\eta^0$, respectively; $\tan\beta=\langle \Phi_2^0
\rangle/\langle \Phi_1^0 \rangle$, and
\begin{eqnarray}
  I_2(x,y,z,w)= \frac{-4 z^2}{z^2-w^2} \int_0^\infty u du
      \left\{\frac{B_1(-u;x,y)-B_1(-u;0,y)}{x^2}\right\}^2
      \left(\frac{z^2}{u+z^2}-\frac{w^2}{u+w^2}\right),
\end{eqnarray}
where $B_1$ is the tensor coefficient function in the Passarino-Veltman's formalism~\cite{passarino-veltman}.
Although the Higgs sector is rather complicated to make
it possible for the electroweak baryogenesis scenario,
the flavor structure is determined only by the combination of 
$h_i^\alpha$ and $m_{N_R^\alpha}$ just as in the Ma model.
The mass matrix has the three loop factor $1/(16\pi^2)^3$
with additional suppression factor by $y_i$.
They are enough to reproduce the neutrino mass scale.
Thus, the electron associated coupling constants $h_e^{1,2}$ and
the scalar coupling $\kappa$ are of ${\cal O}(1)$ for $m_{N_R}^{1,2}\sim {\cal O}(1)$ TeV.
The Yukawa coupling constants $h_i^\alpha$ are hierarchical
as $h_e^{1,2} (\simeq {\cal O}(1)) \gg h_\mu^{1,2} \gg h_\tau^{1,2}$. 

The parameter sets which satisfy the current data from neutrino
oscillation, LFV, relic abundances of DM
and the condition for strongly first order electroweak phase transition 
are studied in Ref.~\cite{aks-prl,aks-prd}. 
To reproduce the neutrino data, the mass of $H^\pm$ should be 100 - 200 GeV.
This is an important prediction of the model. 
In order to avoid the constraint from $b\to s\gamma$, the Yukawa interaction
for the doublet fields takes the form of so-called 
Type-X~\cite{typeX},\footnote{Type-X is referred to as Type-IV in Ref.~\cite{Ref:Barger} and 
Type-I' in Ref.~\cite{grossman}.}
where only one of the
doublets couples to leptons and the rest does to quarks. The physics of
the Type-X two Higgs doublet model (THDM) shows many distinctive features from
the other type of extended Higgs sectors.  For example, $H$ and $A$
decay mainly into $\tau^+\tau^-$ when $\tan\beta \gsim 3$ and
$\sin(\beta-\alpha)\simeq 1$~\cite{typeX}.
There are basically two DM candidates, $\eta^0$ and $N_R^\alpha$. 
The mass of $S^\pm$ is strongly
constrained by the current data and the requirement for strongly
first order phase transition~\cite{aks-prl,aks-prd}.
The coupling constant of $S^+S^-h$ is required to be of ${\cal O}(1)$,
whose indirect effect appears in the quantum correction to the $hhh$ coupling
constant as a large deviation from the SM prediction~\cite{aks-prd,ewbg-thdm2}.
As long as kinematically allowed, $S^\pm$ decays via $S^\pm \to H^\pm \eta^0$ by 100\%.

\section{Phenomenology in radiative seesaw models at the LHC}

The existence of the extra Higgs bosons such as
charged scalar bosons, which are a common feature of radiative
seesaw models, can be tested at the LHC. 
Details of the properties of such extra Higgs bosons
are strongly model dependent, so that we can distinguish
models via detailed measurements of extra Higgs bosons.
In addition, as the (SM-like) Higgs boson $h$ is expected to be
detected, its mass and decay properties are thoroughly measured~\cite{Aad:2009wy}.
The radiative seesaw models with a DM candidate
can also be indirectly tested via the invisible decay of $h$
as long as its branching ratio is more than about $25$\% for $m_h=$ 120 GeV
with ${\cal L}=$ 30 fb$^{-1}$~\cite{LHCinv}.
The phenomenological analyses at the LHC in each model are in the
literature~\cite{Babu:2002uu,AristizabalSierra:2006gb,Nebot:2007bc,kingman-seto,aks-prd,Barbieri:2006dq,CaoMa,typeX,Goh:2009wg,Su:2009fz,Logan:2009uf,Belyaev:2009zd}.
We here review some remarkable features.

At the LHC, $\omega^\pm$ and $k^{\pm\pm}$ in the Zee-Babu model can be
produced in pair, via the Drell-Yan $s$-channel processes 
$q \bar{q} \to \omega^+\omega^-$ and 
$q \bar{q} \to k^{++}k^{--}$. 
The direct detection of $k^{\pm\pm}$ can be a signature for this model.
The $\sigma(q \bar{q} \to k^{++}k^{--})$ is around $0.1$ fb for $m_k\sim 800$ GeV. 
If $k^{\pm\pm}$ mainly decay into $4\mu$ (or $4e$), the number of the signal event 
with ${\cal L}$=100 fb$^{-1}$~\cite{Gunion:1996pq}
 is enough for the discovery for 
$m_{k}\lsim 800$ GeV~\cite{Babu:2002uu}.
The doubly-charged Higgs bosons are however also predicted in the models
with complex triplet scalar fields, $\Delta=(\Delta^{\pm\pm}, \Delta^\pm, \Delta^0)$. 
The gauge coupling $W^\pm \Delta^\pm \Delta^{\mp\mp}$ induces
the single doubly-charged Higgs production 
$q \bar{q'} \to W^{\pm *}\to \Delta^{\pm\pm}\Delta^\mp$, whose cross section
is comparable to that of $q \bar{q} \to \gamma^*, Z^*\to \Delta^{++}\Delta^{--}$~\cite{Akeroyd:2005gt}.
The absence of the gauge coupling
$W^\pm \omega^\pm k^{\mp\mp}$ is an important distinctive
feature of the Zee-Babu model with gauge singlet doubly-charged Higgs bosons 
from the triplet model.
The singly-charged Higgs boson $\omega^\pm$ would be more difficult
to see the signal at the LHC because the final state from $\omega^+\omega^-$
is $\ell^+\ell^-$ plus a missing energy.

%

In the Ma model, if the mass of the DM candidate satisfies $m_{\xi_{r,i}} < m_h/2$,
the SM Higgs boson with $ m_h \lsim 2m_W$
decays mainly into the DM pairs~\cite{Barbieri:2006dq,CaoMa}.
For $m_{\xi_r}=50$ GeV and $m_{\xi_i}=60$ GeV, the branching ratio of the invisible decay 
$h\to \xi_r^0\xi_r^0$ reaches to 70\% around $m_h\sim$ 120 GeV, which can be observed at the LHC.
The $h\to \gamma \gamma$ mode is suppressed by an additional contributions from $\xi^\pm$~\cite{CaoMa}.
In Ref.~\cite{CaoMa}, the discovery potential of the dark scalar doublet is also analyzed by $pp\to \xi_i^0 \xi_r^0$ 
for the benchmark points, $m_{\xi_r}=50$ GeV, $m_{\xi_i}$ = 60 - 80 GeV, and $m_{\xi^\pm}=170$ GeV.
The decay branching ratio of the CP-odd dark scalar, $\xi_i^0 \to Z^*\xi_r^0 \to \ell\bar \ell \xi_r$, is about 0.09 (0.07) 
for $m_{\xi_i}-m_{\xi_r}=$ 10 (30) GeV.
A signature $\ell\ell$ plus a missing energy in the benchmark scenario may be discovered by the optimal cuts.

%
%

In the AKS model, the invisible decay of the SM-like Higgs boson $h\to\eta^0\eta^0$ can also open if
kinematically allowed.
For the typical scenario in Ref.~\cite{aks-prl,aks-prd}, the branching ratio of
the invisible decay can amount to $B(h\to \eta^0\eta^0) \simeq 36$ (34) \% for
$m_\eta=48$ GeV, $m_h=120$ GeV and $\tan\beta$ = 3 (10), 
so it would be testable at the LHC. As the coupling of $h S^+S^-$ is
strong, the partial width of $\Gamma(h\to \gamma\gamma)$ deviates 
from the SM prediction. The lepton specific decays of extra Higgs bosons
$A$, $H$ and $H^\pm$ are discriminative feature of
the Type-X THDM~\cite{Goh:2009wg,Su:2009fz,typeX,Logan:2009uf,Belyaev:2009zd}. 
The dominant decay mode of $H$ ($A$) is
$H\to\tau^+\tau^-$; $B(H (A)\to \tau^+\tau^-) \simeq 1$ for $\tan\beta \gsim 3$. The decay into $\mu^+\mu^-$
is suppressed by a factor of $(m_\mu/m_\tau)^2$.
These neutral bosons can be seen by 
$gg \to h, A, H \to \tau^+\tau^-$ ($\mu^+\mu^-$)~\cite{Belyaev:2009zd,Su:2009fz,typeX}.
The doublet originated charged Higgs boson $H^\pm$ (as well as extra
neutral ones) is as light as 100 - 200 GeV, so that the property of
the Type-X THDM can also be tested by 
$pp \to AH^\pm \to \tau^+\tau^-\tau^\pm \nu_\tau$~\cite{typeX,Logan:2009uf}
and 
$pp \to AH \to 4 \tau$~\cite{typeX,Goh:2009wg}.
On the other hand, the mass of the $Z_2$-odd charged Higgs bosons $S^\pm$
is around 400 GeV in the typical scenario in Ref.~\cite{aks-prl,aks-prd}.
They are produced in pair via the Drell-Yan process,
and decay as $S^+S^-\to H^+H^- \eta^0\eta^0 \to \tau^+\tau^-\nu\nu\eta^0\eta^0$.
The event rate is about 0.5 fb for $m_{S^\pm} = 400$ GeV when $\tan\beta\gsim 2$.
Separation of the $S^+S^-$ signal from the $H^+H^-$ event and also
the SM backgrounds seems to be challenging.


At the LHC, via the physics of extra scalar bosons such as (singly and/or
doubly) charged Higgs bosons and CP-even Higgs bosons, the structure of
the extended Higgs sector can be clarified to some extent.
In addition, the invisible decay of the SM-like Higgs boson and 
the mass spectrum of the extra Higgs bosons 
would give important
indication for a possibility to a radiative seesaw scenario. 
However, although they would  be a strong indication of radiative seesaw models,
one cannot conclude that such Higgs sector is of the radiative seesaw
models. In order to further explore the possibility to such models,
we have to explore the other common
feature of radiative seesaw models, such as the Majorana nature. 
In the next section, we discuss a possibility of testing the Majorana
nature at ILC experiments.

\section{Phenomenology in radiative seesaw models at the ILC}

At the ILC, properties of the Higgs sector can be
measured with much better accuracy than at the LHC, so that
we would be able to reconstruct the Higgs potential in any
extended Higgs sector if kinematically accessible. 
Invisible decays of the Higgs boson can also be tested when the branching ratio
$B(h\to invisible)$ is larger than a few \%~\cite{Schumacher:2003ss}.
Furthermore, the Majorana nature in radiative seesaw models; {\it i.e.,}
the existence of TeV scale right-handed Majorana neutrinos or
that of lepton number violating interaction,
would also be tested at the ILC.

\subsection{Electron-positron collisions}

In the pair production of charged scalar bosons at the $e^+e^-$
collision, which appear in the radiative seesaw models ($\omega^+\omega^-$
 in the Zee-Babu model, $\xi^+\xi^-$ in the Ma model, and $S^+S^-$ (and
 $H^+H^-$) in the AKS model), there are diagrams
of the $t$-channel exchange of left-handed neutrinos
or right-handed neutrinos
in addition to the usual Drell-Yan type $s$-channel diagrams.
The contribution of these $t$-channel diagrams is one of the discriminative
features of radiative seesaw models, and no such contribution enters into
the other extended Higgs models such as the THDM.\footnote{
In the minimal supersymmetric standard model (MSSM) selectron pair production can have similar $t$-channel contributions
   (the Bino exchange). In such a case, the final state would be something
   like a $e^+e^-$
   pair plus a missing energy. 
   Therefore, we can discriminate it from the radiative seesaw models. 
}
These $t$-channel effects show specific dependences on
the center-of-mass energy $\sqrt{s}$ in proportion to $\log s$
in the production cross section, and enhances the production
rates of the signal events for higher values of $\sqrt{s}$.
The final states of produced charged scalar boson pairs
are quite model dependent but with missing energies; 
\begin{eqnarray}
&&  e^+e^- \to \omega^+\omega^- 
  \to \ell_L^+\ell_L^- \underline{\overline{\nu}_{L} \nu_L^{}}\, ,
  \hspace{52mm} [{\rm The~Zee}{\rm-Babu~model}]\\
&&  e^+e^- \to \xi^+\xi^- \to W^{+(\ast)}W^{-(\ast)} \xi_r^0 \xi_r^0 \to
 jjjj(jj\ell_L\underline{\nu_L}) \underline{\xi_r^0 \xi_r^0}\, , 
  \hspace{4.5mm} [{\rm The~Ma~model}]\\
&&  e^+e^- \to S^+S^- \to H^+H^-\eta^0\eta^0 \to \tau^+_R\tau^-_R\underline{\overline{\nu}_L \nu_L \eta^0\eta^0}\, , 
 \hspace{16mm}  [{\rm The~AKS~model}] 
 \end{eqnarray}
where underlined parts in the final states are observed as missing energies.

\vspace{5mm}
\underline{\it The Zee-Babu model}
 
In the Zee-Babu model, the decay branching pattern of $\omega^\pm$ is
determined by the relative magnitudes of the coupling constants
$f_{ij}$. 
As a reference scenario, we take a parameter set
\begin{eqnarray}
&& m_{\omega}=300 {\rm ~GeV}, ~m_{k}=1200 {\rm ~GeV}, ~\mu=800 {\rm ~GeV}, \nonumber\\
&& f_{e\mu}=f_{e\tau}=0.013, ~f_{\mu\tau}=0.027,  \nonumber\\
&& g_{ee} = 0.17, ~g_{\mu\mu} = 1.8, ~g_{\tau\tau}=0.0061,
   ~g_{e\mu} = 5.7\times 10^{-5}, ~g_{e\tau} = 0.011, ~g_{\mu\tau}=-0.081,~~ 
   \label{eq:para-babu}
\end{eqnarray}
which satisfies the neutrino data for the normal mass hierarchy.
In this scenario, the rate in final states
$\ell^+\ell^-=e^+e^-$, $e^\pm\mu^\mp$, $e^\pm \tau^\mp$, $\mu^+\mu^-$,
$\mu^\pm\tau^\mp$, $\tau^+\tau^-$
is given by $2$, $13$, $13$, $19$, $36$, $19$, respectively.
In Fig.~2,  the differential cross
section $d\sigma/d\cos\theta_\mu$ for the signal
$e^+e^- \to \omega^+\omega^-\to \mu^- \tau^+$ (+missing energy)
is shown for $\sqrt{s}=1$ TeV as a function of $\cos\theta_\mu$,
where $\theta_\mu$ is the angle 
between the outgoing muon and the beam axis. 
The main background comes from the $W$ pair production, which is
also plotted.
The angle cut ({\it e.g.} $\cos\theta_\mu < -0.5$) improves the ratio of the signal and 
the background.\footnote{Although in this Letter we mainly discuss the case where
$m_{k}$ is at the TeV scale, we just comment on the case of lighter 
$k^{\pm\pm}$.  
In such case, the pair production of $k^{++}k^{--}$ can be a clear
signature of this model, whose signal is the like-sign dilepton pairs
with opposite direction~\cite{ZeeBabu,Babu:2002uu}. 
}
\begin{figure}[t]
\includegraphics[width=8cm,angle=0]{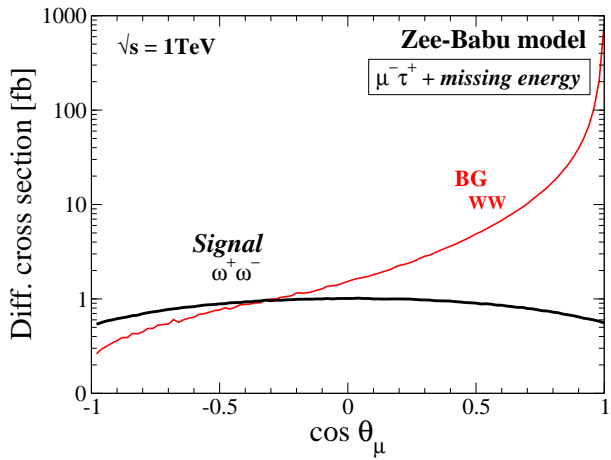}
  \caption{
 The differential cross section of $e^+e^-\to \omega^+\omega^-\to
 \mu^-\tau^+$(+missing energy) as a function of 
 the angle 
 between the outgoing muon and the beam axis
 in the Zee-Babu model for $\sqrt{s}=1$ TeV.
 The rate of $\mu^-\tau^+$(+missing energy) from main background $e^+e^-\to W^+W^-$ is also shown.
 }
  \label{fig:babu}
\end{figure}

At current and future LFV experiments, 
the coupling constants $f_{ij}$ and $g_{ij}$ can be further tested
via the LFV rare decays such as $\ell\to \ell' \gamma$ and $\ell\to \ell'\ell'\ell''$. 
The same operators as in $\ell^- \to \ell'^{\mp} e^-e^{\pm}$
would also be tested directly at the ILC via $e^{\pm}e^- \to \ell^{-} \ell^{'\pm}$.
In the scenario in Eq.~(\ref{eq:para-babu}), we estimate that
$\sigma(e^+e^-\to \mu^\pm \tau^\mp$) $\sim 5$ fb for $\sqrt{s}=1$
TeV. When we take $g_{ee}=0.4$, $g_{e\tau}=0.01$, $m_k=\mu=1.2$ TeV and
$m_\omega=400$ GeV, which also satisfy all the current data, we obtain
$\sigma(e^+e^-\to \tau^\pm e^\mp) \sim 0.76$ (1.7) fb for
$\sqrt{s}=500$ GeV (1 TeV).  

\begin{figure}[t]
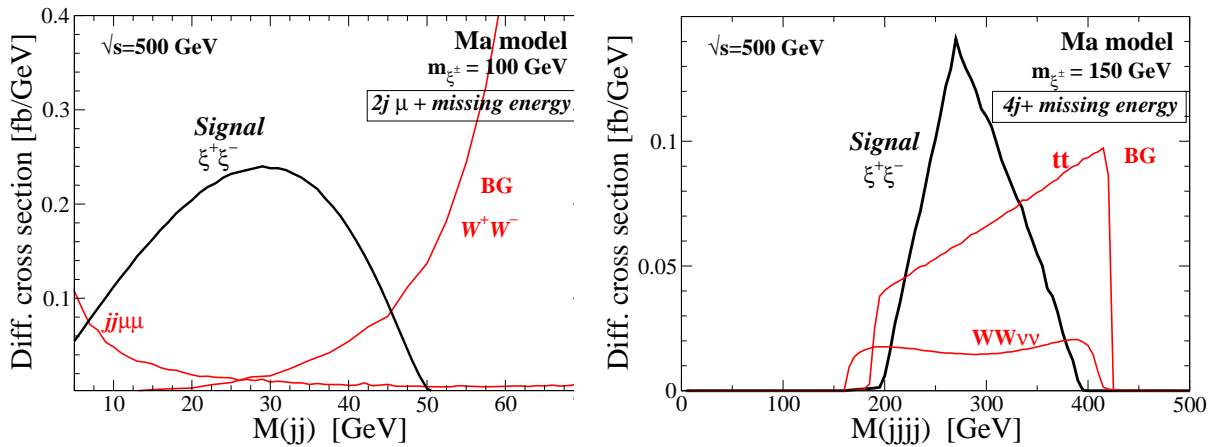

\includegraphics[width=7.8cm,angle=0]{Mjj.eps}
\includegraphics[width=8cm,angle=0]{M.eps}
  \caption{
The jets invariant mass distributions of the production rates of the signal in the Ma model at $\sqrt{s}=500$ GeV. 
{\it left} : The di-jet invariant mass $M(jj)$  distribution of the signal $e^+e^-\to \xi^+\xi^-  \to jj\mu\nu\xi^0_r\xi_r^0$ for $m_{\xi^\pm}=100$ GeV. {\it right} : $M(jjjj)$ distribution of $e^+e^- \to \xi^+\xi^- \to W^+W^- \xi_r^0 \xi_r^0 \to jjjj \xi_r^0$ for $m_{\xi^\pm}=150$ GeV. In addition to the rate from the signal process, those for main backgrounds are also shown.
 }
  \label{fig:ma}
\end{figure}

\vspace{5mm}
\underline{\it The Ma model}

In the Ma model, the coupling constants $\hat{h}_{e}^\alpha$ ($\alpha=1,2$)
\footnote{Here we consider the minimal case of two generations for the right-handed neutrino.}
are strongly constrained from neutrino data and LFV data. 
As a typical choice of parameters, we consider
\footnote{
The relatively large mass difference between $\xi^\pm$ and $\xi_{r,i}^0$
implies a significant deviation from the custodial symmetry in the
Higgs sector, which affects the allowed mass $m_h$ of the SM like Higgs
boson $h$. The larger $m_h$ is favored for larger mass difference of $m_{\xi^\pm}-m_{\xi_{r,i}}$.
}
\begin{eqnarray}
&&  m_{\xi_r}=50 {\rm ~GeV},  ~m_{\xi_i}=60 {\rm ~GeV},
 ~m_{\xi^\pm}\sim100 {\rm ~GeV}, 
 ~m_{N_R^1}  = m_{N_R^2}  
 = 3 {\rm ~TeV},
  \nonumber \\
&&  
\lambda_5 = -1.8 \times 10^{-2} , ~~\hat{h}_e^\alpha, \hat{h}_\mu^\alpha, \hat{h}_\tau^\alpha
\sim 10^{-5},
 \label{eq:para-ma}
\end{eqnarray}
in which the normal neutrino mass hierarchy is realized.
Because $\hat{h}_e^\alpha$ are very small for a TeV scale $m_{N_R^\alpha}$,
the contribution of the $t$-channel diagrams to the signal $e^+e^-\to\xi^+\xi^-$ 
is much smaller than that from Drell-Yan type diagrams.
For most of the possible values of $\hat{h}_\ell^\alpha$ and $m_{N_R^\alpha}^{}$
  which satisfy the LFV and the neutrino data, the contribution of the
  $t$-channel diagrams is negligible.
The production cross section of a charged Higgs pair $\xi^+\xi^-$
is therefore similar to that in the usual THDM:
about 92 (10) fb for $m_{\xi^\pm}=100$ (150) GeV at $\sqrt{s}=$ 500 GeV.
The produced $\xi^\pm$ decay into $W^{\pm (*)} \xi_{r,i}^0$.
\footnote{
The $\xi_r^0 \xi_i^0$ production can also
be interesting. The final state should be two jets (or dilepton) plus a missing energy.
The cross section for $e^+e^-\to \xi_r^0 \xi_i^0 \to \xi_r^0\xi_r^0 jj$ is about 40 fb at $\sqrt{s}=500$ GeV.
}

 In Fig.~3 ({\it left}), we show the invariant
mass distribution of the di-jet $jj$ of the production cross section of the
signal, $e^+e^-\to \xi^+\xi^- \to W^{+ \ast} W^{-\ast} \xi_r^0 \xi_r^0 \to jj\mu\nu\xi^0_r\xi_r^0$
for $m_{\xi^\pm}=100$ GeV.
The main backgrounds come from $WW$.
The $jj\mu\mu$ events from $ZZ$, $\gamma\gamma$, and $Z\gamma$
can also be the backgrounds.
A factor of 0.1 is multiplied to the rate of the $jj\mu\mu$ backgrounds
for the miss-identification probability of a muon. 
The signal is significant around $M(jj)\sim$30 GeV. 
The invariant mass cut (such as 15 GeV$< M(jj) <$ 40 GeV) is effective 
to reduce the backgrounds.
For the numerical evaluation, we have used a package CalcHEP 2.5.4~\cite{calchep}.

For $m_{\xi^\pm} > m_W+m_{\xi_r}$, on the other hand, 
the signal $W^+W^-\xi^0_r\xi^0_r$ can be measured by
detecting the events of four jets with a missing energy. The main background
comes from $W^+W^-\nu\nu$ and $t\overline{t}$. 
By the invariant
mass cuts of two-jet pairs at the $W$ boson mass, the biggest background
from $WW$ can be eliminated. In Fig.~3 ({\it right}), we show the invariant
mass distribution of $jjjj$ of the production cross sections of the
signal and the backgrounds without any cut.
A factor of 0.1 is multiplied to the rate of $tt$ background, by which
the probability of the lepton from a $W$ that escapes from detection is
approximately taken into account.  
The signal is already significant. The invariant mass cut
($M(jjjj)< 300$ GeV) gives an improvement for the signal/background ratio.

\begin{figure}[t]
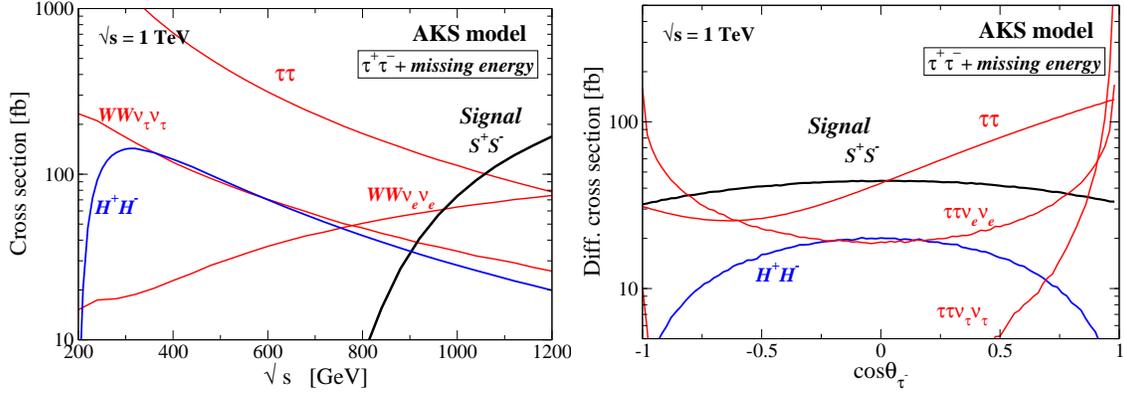

\includegraphics[width=7.5cm,angle=0]{roots.eps}
\includegraphics[width=7.2cm,angle=0]{AKS_angle.eps}
  \caption{
{\it left} : The cross sections of the signal, $e^+e^-\to S^+S^- \to
\tau^+\tau^-$ (+ missing energy), in the AKS model as a function of 
the collision energy $\sqrt{s}$.
{\it right} : The differential cross section of the signal for $\sqrt{s}=$ 1 TeV as a function of the
angle of the direction of the outgoing $\tau^-$ and the beam axis of incident electrons.
 In addition to the rate from the signal, those from backgrounds 
 such as $\tau^+\tau^-$, $\tau^+\tau^- \overline{\nu} \nu$ and
 $H^+H^-$ are also shown.
 }
  \label{fig:aks}
\end{figure}

\vspace{5mm}
\underline{\it The AKS model}

For the AKS model, we take a typical successful scenario for the neutrino data
with the the normal mass hierarchy, 
the LFV data and the DM data as well as the condition for strongly first order phase
transition~\cite{aks-prl,aks-prd}; 
\begin{eqnarray}
&&m_{\eta}=50 {\rm~GeV}, ~m_{H^\pm}=100 {\rm~GeV}, ~m_{S^\pm}=400 {\rm~GeV}, 
~m_{N_R^1}^{}=m_{N_R^2}^{}=3 {\rm~TeV}, \nonumber \\
&&h_{e}^1=h_{e}^2=2 \gg h_{\mu}^1,h_{\mu}^2  \gg h_{\tau}^1,h_{\tau}^2, 
~\kappa \sim{\cal O}(1),
~\sin(\beta -\alpha)=1,~\tan\beta=10\, .
\label{eq:para-AKS}
\end{eqnarray}
Because $h_e^{1,2}\sim$ ${\cal O}$(1), the contribution
from the $t$-channel $N_R^\alpha$ exchange diagrams
to the production cross section of $S^+S^-$ dominate that from the 
Drell-Yan diagrams~\cite{aks-prd}. The cross section is
about 87 fb for $m_{S^\pm} = 400$ GeV at $\sqrt{s}=1$ TeV. 
As the decay branching ratio of $S^\pm \to H^\pm \eta^0$ is 100\% and
that of $H^\pm \to \tau^\pm \nu$ is also almost 100\% because of the
Type-X THDM interaction for $\tan\beta = 10$, the final state of the
signal is $\tau^+\tau^- \overline{\nu} \nu \eta^0\eta^0$ with almost the same
 rate as the parent $S^+S^-$ production.
The main SM backgrounds are
$\tau^+\tau^-$ and $\tau^+\tau^- \overline{\nu}\nu$.
The pair production of the doublet like charged Higgs boson $H^+H^-$
can also be the background. 
As the signal rate dominantly comes from the $t$-channel diagram,
it becomes larger for larger $\sqrt{s}$, while
the main backgrounds except for $\tau\tau\nu_e\nu_e$
are smaller because they are dominantly $s$-channel processes (Fig.~4 ({\it left})). 
At $\sqrt{s}=1$ TeV, the rate of the signal without cut
is already large enough as compared to those of the backgrounds.
It is expected that making appropriate kinematic cuts  will
improve the signal background ratio to a considerable extent.
The $\sqrt{s}$ scan will help us to confirm that the signal rate
comes from the $t$-channel diagrams. Fig.4 ({\it right}) shows the differential cross 
section of the signal at $\sqrt{s}=$ 1 TeV as a function of $\cos\theta_{\tau^-}$, where
$\theta_{\tau^-}$ is the angle between the direction of the outgoing $\tau^-$ and 
the beam axis of incident electrons.
The distribution of the background from $\tau\tau$ is 
asymmetric, so that the angle cut for larger $\cos\theta_{\tau^-}$ reduces the backgrounds.

\subsection{Electron-electron collisions}

As already stated, the ILC has a further advantage to test radiative seesaw models via the
experiment at the $e^-e^-$ collision option, where dimension five
operator of $e^-e^- \phi^+\phi^+$, which is 
the sub-diagram of the 
loop diagrams for neutrino mass matrix. 
This direct test of the dimension five operator is essential
to identify the radiative seesaw models.

The Majorana nature in the Zee-Babu model is in the lepton number violating
coupling constant $\mu$ of $k^{++} \omega^-\omega^-$, which generates the dimension five
operator of $e^-e^- \omega^+ \omega^+$ at the tree level
via the $s$-channel $k^{--}$ exchange diagram.
The cross section of $e^-e^- \to \omega^-\omega^-$ is given by
\begin{eqnarray}
 \sigma(e^-e^- \to \omega^-\omega^-) = \frac{1}{8 \pi}
  \sqrt{1-\frac{4 m_{\omega}^2}{s}} 
  \frac{\mu^2 g_{ee}^2}{(s-m_{k}^2)^2+m_{k}^2 \Gamma_{k}^2}, 
 \end{eqnarray}
 where the total width $\Gamma_k$ of $k^{\pm\pm}$ is computed as about 168 GeV in our scenario
 in Eq.(\ref{eq:para-babu}).
On the other hand, in the Ma model and the AKS model, the operator
comes from the $t$-channel right-handed neutrino exchange diagram.
The cross section is evaluated as
\begin{eqnarray}
  \sigma(e^-e^- \to \phi^-\phi^-) =
      \int_{t_{\rm min}}^{t_{\rm max}} dt \frac{1}{128\pi s}
           \left|\sum_{\alpha=1}^n  (|c^\alpha|^2
           m_{N_R^\alpha}) \left(\frac{1}{t-m_{N_R^\alpha}^2}+\frac{1}{u-m_{N_R^\alpha}^2}\right) \right|^2,
 \end{eqnarray}
 where $n$ is the number of generation of right-handed neutrinos,
 $\phi^-$ represents the $Z_2$-odd charged scalar boson
 $\xi^-$ in the Ma model and $S^-$ in the AKS model. The constants
 $c^\alpha$ represent $\hat{h}_e^\alpha$ or $h_e^\alpha$ in the Ma
 model or the AKS model, respectively.
We note that due to the Majorana nature of the $t$-channel diagram, we
 obtain much larger cross section in the $e^-e^-$ collision than in the
 $e^+e^-$ collision in each model assuming the same collision energy. 

\begin{figure}[t]
\includegraphics[width=8cm]{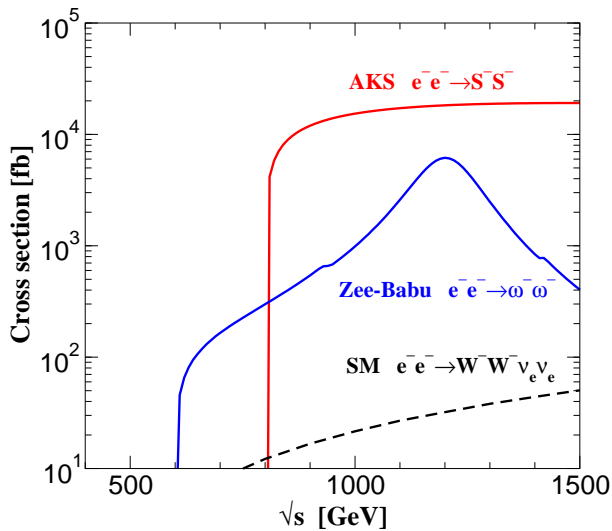}
  \caption{The cross sections of like-sign charged Higgs pair
 productions in the Zee-Babu model ($\omega^- \omega^-$)
 and in the AKS model ($S^-S^-$) are shown as a function of the collision energy
 $\sqrt{s}$. The parameters in the Zee-Babu and the AKS model are taken as in
 Eq.~(\ref{eq:para-babu}) and Eq.~(\ref{eq:para-AKS}), respectively. 
 }
  \label{fig:ee}
\end{figure}

The mass matrix of left-handed neutrinos is generated at the one, two and three loop levels
in the Ma model, the Zee-Babu model and the AKS model, respectively.
Therefore, the coupling constants can be basically hierarchical among the models, so are
the cross sections.
For the typical scenarios in these models,
the cross sections are shown in Fig~5. 
The rate of $\omega^-\omega^-$ in the Zee-Babu model can be larger than
several times 100 fb for 800 GeV $\lsim \sqrt{s} \lsim 1.5$ TeV.
It becomes maximal (several times pb) at $\sqrt{s}\sim m_k$,
and above that asymptotically reduces by $1/s$.
The maximal value of the cross section is sensitive to the value of
$g_{ee}$ and $\mu$. In the parameter sets where these coupling constants
are smaller the cross section becomes smaller.
The signal should be like-sign dilepton with a missing energy.
On the other hand, in the Ma model, production cross sections of $e^-e^- \to
\xi^-\xi^-$ are smaller than $10^{-4}$ fb because the coupling
constants $\hat{h}_i^\alpha$ are very small in the parameters in Eq.~(\ref{eq:para-ma}).
Allowing some fine tuning, $\hat{h}_i^\alpha$ may be at most 0.01 for heavier $N_R^\alpha$.
In any case, the cross section of $e^-e^- \to \xi^-\xi^-$ is smaller than
$10^{-3}$ fb. Hence, most of the successful scenarios in the Ma model
the process $e^-e^- \to \xi^-\xi^-$ is difficult to be seen. 
In the AKS model, the cross section of $e^-e^-\to S^-S^-$ 
is large, and its value amounts to about 15 pb at $\sqrt{s}=1$ TeV in the
scenario given in Eq.~(\ref{eq:para-AKS}).
Above the threshold, the magnitude of
the cross sections are not sensitive to $\sqrt{s}$,
so that even if $m_{S^\pm}^{}$ would be at the TeV scale, 
we might be able to test it
at future multi-TeV linear colliders, such as the Compact Linear Collider~\cite{CLIC}.
Because $B(S^\pm \to \eta^0 H^\pm) \simeq B(H^\pm \to \tau^\pm \nu) \simeq 100$~\%,
the signal should be $\tau^-\tau^-\nu\nu\eta\eta$ with almost the same
rate as long as $m_{S^\pm}^{} < m_{N_R^\alpha}$.  

The background mainly comes from $W^-W^-\nu_e\nu_e$, 
and the cross section is about 2.3 fb (22 fb) for $\sqrt{s}=500$
GeV (1 TeV). The branching ratio for the leptonic decay of $W$ bosons is 30\%, so that the
rate of the final state $\ell\ell'\nu\nu\nu\nu$ is at most 2 fb or less.
Therefore, the signal in the AKS model and in the Zee-Babu model can be seen.

Apart from the TeV-scale radiative seesaw models, there are many models with
lepton number violating interactions or right-handed Majorana
neutrinos.
Atwood et al. have discussed the signature of heavy Majorana neutrinos
in the model without $Z_2$ symmetry via charged Higgs pair production at
$e^+e^-$ and $e^-e^-$ collisions~\cite{atwood}.\footnote{Recently, charged Higgs pair production 
at $e^-e^-$ collider is studied in a specific model with three TeV-scale right-handed neutrinos and 
four Higgs doublets in Ref.~\cite{grimus}.}
In supersymmetric models, Majorana particles also appear, and their
effects also give similar $t$-channel contributions to the above models in the slepton pair production through
the gauge couplings, $e^-e^- \to \tilde{e}^- \tilde{e}^-$, 
whose cross section is of ${\cal O}$(100) fb.
The final state would be $e^-e^- \chi^0\chi^0$ for example.

The $e^-e^-$ collision experiment is useful to test
the Majorana nature of radiative seesaw models such as the Zee-Babu
model and the AKS model via like-sign pair production of charged scalar bosons. 
The cross section can be significant and hierarchical among these models.
The signal can be observed as a model dependent final state, by which
we can discriminate the models.
Although in this Letter we did not explicitly discuss the KNT model that
 also contains TeV-scale right-handed neutrinos $N_R^\alpha$, we found that the cross section of
 $e^-e^- \to S_2^-S_2^-$ ($S_2^-$ is the $Z_2$-odd isosinglet charged scalar boson) is very small because the coupling
 constants for $\overline{e}_R N_R^\alpha S_2^-$ ($\alpha=1,2$) are
 tiny~\cite{kingman-seto}.

\section{Conclusion}

 We have discussed 
 general features of TeV-scale radiative seesaw models. They are 
characterized by an extended scalar (Higgs) sector and the Majorana nature.
We have mainly discussed the concrete models with neutrino mass generation at
one-loop (the Ma model), two-loop (the Zee-Babu model), and at three-loop 
(the model in Ref.~\cite{aks-prl}).
Various phenomenological aspects of these models have been discussed 
 especially in experiments at the LHC and at the ILC.
 We have found that, while the extended Higgs sector can be explored at the LHC, 
 the Majorana nature of the models can directly
be tested at the ILC via the pair production of the charged scalar bosons at
the electron-positron and electron-electron
 collision experiments.
The detailed realistic simulation has to be done elsewhere. 

\vspace{5mm}
\noindent
{\it Acknowledgments}

We like to thank Ernest Ma for useful discussions. 
The work of SK was supported, in part, by Grant-in-Aid for scientific
research (C), Japan Society for the Promotion of Science (JSPS), No.~19540277.

\end{document}